# Band Structure Engineering in Highly Crystalline Organic Semiconductors

Shu-Jen Wang*, Sebastian Hutsch, Felix Talnack, Marielle Deconinck, Shiyu Huang, Zongbao Zhang, Hans Kleemann, Yana Vaynzof, Stefan C. B. Mannsfeld, Frank Ortmann* and Karl Leo*

Dr. S-J. Wang, Ms. M. Deconinck, Ms. S. Huang, Mr. Z. Zhang, Dr. H. Kleemann, Prof. Y. Vaynzof, Prof. K. Leo
Dresden Integrated Center for Applied Physics and Photonic Materials (IAPP), Technische Universität Dresden, 01069, Dresden, Germany
Email: shu-jen.wang@tu-dresden.de; frank.ortmann@tum.de; karl.leo@tu-dresden.de

Dr. S. Hutsch, Prof. F. Ortmann
Technische Universität München, Department of Chemistry, Lichtenbergstr. 4, 85748 Garching b. München, Germany

Mr. F. Talnack, Ms. M. Deconinck, Mr. Z. Zhang, Prof. Y. Vaynzof, Prof. S. C. B. Mannsfeld, Prof. K. Leo
Center for Advancing Electronics Dresden (cfaed), Technische Universität Dresden, Helmholtz Str. 18, 01069 Dresden, Germany

Ms. M. Deconinck, Mr. Z. Zhang, Prof. Y. Vaynzof
Leibniz Institute for Solid State and Materials Research, IFW, Helmholtzstraße 20, 01069 Dresden, Germany

**Abstract**

Blending of semiconductors for controlling the energy levels (band structure engineering) is an important technique, in particular, for optoelectronic applications. The underlying physics is the delocalized Bloch states, which average over the potential landscape of the blend. For organic semiconductors, it has been shown that two quite different effects, the dielectric constant and electrostatic interaction between molecules, can be used to tune the energy gap and ionization energy of disordered and weakly crystalline organic semiconductor blends. It is so far not known whether the electronic delocalization in organic crystals with large bandwidths can contribute to the energy structure engineering of the blend in a way similar to that in inorganic semiconductors. Here, we investigate the growth of highly ordered organic thin-film blends with a similar chemical structure and show the effect of band structure engineering by spectroscopic methods. We rationalize the experimental results with comprehensive theoretical simulations, showing that the delocalization is a significant effect. Our work paves the way for engineering the band structure of highly ordered organic semiconductor thin films that can be tailored for the desired optoelectronic device application.

**Introduction**

Organic semiconductors are dominating the commercial smartphone display market and are rapidly enabling emerging device applications including aesthetic solar windows and photodetectors due to their excellent optical properties, ease of processing and low cost[1-2]. The ability to fine-tune the energy levels and band gaps of the semiconductors is of crucial importance for modern electronics to enable efficient electronic components. In crystalline inorganic semiconductors such as GaAs, band structures can be tuned continuously by blending materials with different energy levels through delocalized Bloch states that averaged over different compositions of the alloys[3-5]. Organic semiconductors, on the other hand, are typically characterized by their localized polarons, which makes energy structure tuning through electronic interaction impractical[6]. However, energy level and energy gap tuning have been demonstrated in organic semiconductors upon blending different molecules through a different physical origin involving electrostatic interaction and dielectric constant differences[7-9]. Research efforts dedicated to energy structure engineering in organic semiconductors thus far have focused on disordered and weakly crystalline films, whose charge carriers can be well defined by the localized polaron concept. In contrast, highly ordered organic crystals - such as pentacene and rubrene - exhibit a bandwidth on the order of a few hundred meV, meaning that the charge carriers are delocalized over several molecules[10-12]. Therefore, it is interesting to explore energy structure engineering in such systems. To achieve energy structure tuning in systems with strong electronic coupling, it might be necessary to select molecules of similar size and structure to allow the formation of intermixed crystals, but with different energy gaps.

This is, however, challenging since molecules with different energy gaps typically have different molecular size and structure.

High quality organic single crystals are typically prepared by physical vapor transport (PVT) method where molecules are sublimed along a glass furnace and individual crystals are formed along the tube furnace[13]. Even though the single crystals produced by PVT are of high quality, transferring such crystals for device integration is tedious and complex. In addition, the size, thickness and composition of the PVT grown crystals cannot be accurately controlled. Hence, it is difficult to scale-up PVT grown crystals toward industrially relevant applications. On the other hand, some organic semiconductors with certain thermal and molecular properties show the ability to form highly ordered thin film crystals upon annealing of amorphous evaporated films as well as epitaxial growth of molecules deposited on top of already formed crystals[14,15]. Rubrene has been an archetypal material of crystallizable organic semiconductors that has been extensively investigated due to its superior charge and exciton transport properties[16-20]. It has been demonstrated that a small number of dopants can be incorporated into crystalline rubrene films to modulate their electrical and functional properties[21,22], which allows the construction of multi-junction vertical devices such as solar cells, light-emitting diodes and organic bipolar junction transistors harnessing the benefits of excellent charge and exciton transport of the crystalline rubrene[23-25].

In this work, we investigate the energy level engineering of crystalline rubrene blends with molecules with similar chemical structures. We provide a detailed study of the blend and heterostructures by X-ray diffraction, atomic force microscopy (AFM) and ultraviolet photoelectron spectroscopy (UPS). Our experimental observations of energy structure engineering are rationalized by theoretical simulation of the molecular properties of the blend films.

**Results and Discussion**

**Crystalline organic semiconductor blend films and heterostructures.**

We choose rubrene (RUB) as the material to initiate this study since it is a model organic semiconductor with strong intermolecular interactions and a large bandwidth in crystalline form (0.4 eV energy dispersion of the valence states in single crystal rubrene[10]). In addition, amorphous films of RUB can be transformed into polycrystalline films with a large domain size of a few hundreds of microns by a simple thermal annealing method, as well as the ability to grow subsequent rubrene layers epitaxially (homoepitaxy) by depositing molecules on to the already formed crystalline films. To maximize the possibility of commensurate epitaxy and well distributed blending at a molecular level, two molecules with similar chemical structure and size as RUB are selected that have either part [5,12-Diphenyltetracene (DPT)] or all [tetracene (TET)] of the phenyl rings removed from the core of the rubrene molecule (shown in Fig. 1a). Moreover, these two molecules have different singlet ($S_1$) and triplet ($T_1$) energies (DPT [$S_1$ = 2.54 eV ($T_1$ = 1.23 eV)] and TET [$S_1$ = 2.75 eV ($T_1$ = 1.29 eV)]) in comparison to rubrene [$S_1$ = 2.34 eV ($T_1$ = 1.11 eV)], which makes them promising candidates for energy structure engineering[26].

It is worth noting that neither pure DPT or TET films can form crystalline films by means of annealing in which, there appears to be no change to the DPT films after annealing at low temperature (~100 °C) and the films de-wets from the substrate surface at higher annealing temperature of approximately 160 °C (Fig. S1). This finding is likely related to the lower amount/ lack of phenyl rings, that restricts the molecular motion on the substrate surface and prohibits the formation of large crystallite domains upon annealing[27]. The situation changes when a blended film of DPT and RUB is annealed at 160 °C resulting in large crystalline domains on the order of a few hundred microns (Fig. 1b-d). The crystal growth mode from the nucleation centers changes from round-shape to triangular-shape with increasing DPT blending content. The morphology of the blended crystal shows step-like growth pattern with a low surface roughness (1.79 nm RMS) (Fig. S2). It is interesting to note that the amount of DPT molecules that can be included in the RUB film without compromising the crystallization process is significantly higher than for other dopant molecules where doping concentration in excess of 2 wt. % would suppress the RUB crystals formation upon annealing[21,22]. Furthermore, only the orthorhombic platelet polymorph of RUB seems to be able to accommodate the DPT molecules upon blending as annealing of DPT:RUB films at lower temperature - where the formation of triclinic RUB is expected - did not show any crystallization. Similarly, blend films of TET:RUB did not show any crystallization upon annealing for both orthorhombic and triclinic phases of RUB.

We next turn to another method of blending and creating heterostructures through epitaxial growth. In this approach, we use the RUB's ability in sustaining homoepitaxy with the crystal seed layer already formed and introduce blending and heterostructures by depositing onto a RUB thin film crystal template. Fig. S3 shows the polarized optical images of the blend film and heterostructures deposited on top of orthorhombic RUB crystals. It is clear that pure TET film exhibits grainy morphology covering the entire surface of the film while the blend TET:RUB (1:1) film shows micro-sized grains that appear non-continuous across the film. On the other hand, pure DPT film exhibit a similar grainy morphology, albeit with smaller grains in comparison to TET film, whereas the blend DPT:RUB (1:1) film shows no distinct features similar to plain orthorhombic RUB crystals. AFM was used to obtain a deeper insight into the surface morphology of the films to better understand the molecular growth and the AFM images are shown in Fig. 2. The TET film consists of individual grains of around half a micron in size and the roughness of the film is 5.51 nm RMS (Fig. 2d). Such polycrystalline nature of the TET film is expected and observed in thermally evaporated layers as TET molecules tend to agglomerate and form a micro-crystalline morphology[27]. The blend TET:RUB (1:1) film becomes considerably rougher (20.82 nm RMS) in comparison to the pure TET film and there appears to be separated grains of TET surrounded by RUB terrace-like features. In contrast, the DPT film shows columnar morphology with a surface roughness of 5.7 nm RMS (Fig. 2b) whereas the blend DPT:RUB (1:1) film becomes smoother (1.46 nm RMS) with fine topography features (Fig. 2a).

We used grazing-incidence wide-angle X-ray scattering (GIWAXS) to investigate the crystal structure of the blend and heterostructure films (Fig. 3). For the pure TET films and blend films with RUB, the diffraction pattern of the orthorhombic phase of RUB is observed. In addition to the diffraction pattern of RUB, several additional peaks are visible in the GIWAXS patterns, which are attributed to the TET film. They are visible in the films blended with RUB as well as the pure TET layer on top of RUB, indicating that TET forms separate crystals during the growth of the thin film. The peaks of TET in the diffraction pattern show arc-like characteristics, typical for thick crystalline thin films where the crystallites are disordered to some extent in their orientation towards the substrate plane, as visible in the AFM image in Figure 2. The situation is different for the DPT film and its blend with RUB where no additional diffraction peaks or rings appear other than the orthorhombic RUB peaks in the GIWAXS pattern. The same holds for the DPT:RUB blend that are mixed in the seed layer before annealing (Fig. S4). Our results indicate the RUB and DPT molecules intermix well in a blend crystalline film.

Next, we examine the electronic structure of the blend films using ultraviolet photoemission spectroscopy. Fig. 4 shows the ionization potential (IP) and work function (WF) probed by UPS of the blended film as a function of the blending ratio. There is a continuous shift in ionization potential for the DPT:RUB blend towards a higher value as the blending ratio is increased, whereas no clear dependency can be observed for the TET:RUB blend (Fig. S5b). In addition, the work function appears to be weakly dependent on the blending ratio for both DPT:RUB and TET:RUB blends implying there is no doping or charge transfer in the blend films. The optical absorption and emission of the blend films show a generally expected behavior, where with increasing blending ratio the spectral contribution of the specific molecular component is increased (see Fig. 5). It is worth noting that the RUB absorption peak at around 530 nm is suppressed upon blending with DPT molecules, which indicates a modification of the optical energy gap of the blend layer.

We further investigate the charge transport properties of the crystalline blend films by using a hole-only diode configuration where the blend films are sandwiched between two p-type doped RUB contacts enabling a symmetric device (Fig. 6a). Fig. 6b shows the current density-voltage characteristics of the hole-only diodes based on the blend films. It can be clearly observed that the threshold voltage (i.e. the transition from linear regime to super-linear regime) increases with increasing blending ratio which is characteristics of increased energetic barrier consistent with the UPS results. However, the slope of the curve in the super-linear regime remains high (around 5-6) at high voltages probably due to the non-ideal charge injection and DPT acting as a charge trap making it difficult to extract the carrier mobility from the space charged limited regime by the $V^2$ dependence. In any case, by examining the magnitude of current density vs voltage, the carrier mobility of the blend film can be estimated to be on the order of 0.1 cm$^2$/Vs despite the device non-ideality[21].

To rationalize the observed shift of the IP in dependence of the blending ratio in the DPT:RUB blend we performed simulations of the electronic density of states (DOS) of the blended systems. Therefore, the DOS for the parametrized electronic tight-binding Hamiltonian

$$H = \sum_M \varepsilon_M + \Delta\varepsilon_M + \sum_{M \neq N} \varepsilon_{MN} + \Delta\varepsilon_{MN} \qquad \text{(Eq. 1)}$$

is calculated for different blending ratios. Here, $\varepsilon_M$ are the onsite-energies (corresponding to the IP of the molecules in gas-phase calculated with density functional theory (DFT)) and $\varepsilon_{MN}$ are the transfer integrals between the molecular orbitals of interest, which in the case of the IP are the highest occupied molecular orbitals (HOMO). $\Delta\varepsilon_M$ and $\Delta\varepsilon_{MN}$ are the respective energetic disorders modelled by a Gaussian distribution with disorder strengths $\sigma_M$ and $\sigma_{MN}$, respectively (the exact values are stated further below). Different blending ratios are obtained by changing the amount of sites $M$ corresponding to DPT and RUB, i.e. the sites with onsite energy $\varepsilon_M^{DPT}$ and $\varepsilon_M^{RUB}$. These energies were calculated with DFT as the energetic difference between the charged and neutral molecule[28] and amount to $\varepsilon_M^{DPT} = -6.24 eV$ and $\varepsilon_M^{RUB} = -6.05 eV$. That is, we find a difference of roughly 200 meV in the gas-phase ionization energy due to the different amount of phenyl rings of DPT and RUB. To calculate the DOS of the blended system, as a first guess we assume that all blends and pure DPT crystallize in the ideal orthorhombic crystal structure of RUB. That is, we take the respective (narrowed) transfer integrals $\varepsilon_{MN}$ of orthorhombic rubrene (with the largest unnarrowed being $\varepsilon_{MN}^{max} \approx 104 meV$) and the disorders $\Delta\varepsilon_M\ N(\varepsilon_M, (30 meV)^2)$ and $\Delta\varepsilon_{MN}^{max}\ N(\varepsilon_{MN}^{max}, (43 meV)^2)$ solely stem from the inevitable vibrational disorder (cf. paper[29] for the methodology). The resulting density of states in dependence of the blending ratio are shown in Fig. S5c (all DOS were shifted by the same amount to match the experimental energies). We indeed observe a continuous shift of the weight of the DOS to higher energy with increasing DPT content similar to the experiment. Though, in contrast to experiment, we observe a multi-peak structure that stems from the almost clean band structure since we utilized parameters of ideal orthorhombic RUB. In reality, the pure DPT and blended systems will be more disordered due to the asymmetry of the DPT molecules compared to RUB. Since this complex disorder is not accessible in simulations, we model the increased disorder effectively by reducing the maximum transfer integrals to $\varepsilon_{MN}^{max} \approx 60 meV$ and increase the local disorder $\Delta\varepsilon_M$ for pure DPT by $\sigma_M^{dis} \approx 100 meV$ and for the blended systems by $\sigma_M^{dis} \approx 180 meV$. The resulting DOS are found in Fig. 4b. The distributions match qualitatively to the experimental spectra derived from the UPS peak (ref. Fig 4a,b) and we indeed observe a continuous shift of the peak maxima with blending ratio, which is compared to experimental values in Fig. 4c,d. We thus find that the different electronic properties of RUB and DPT can lead to an energetic shift of the bands and DOS in the blended systems in dependence of the mixing ratio.

**Conclusions**

In summary, we investigated the growth of highly ordered molecular blends and heterostructures based on rubrene and a tetracene derivative. We further show the band structure engineering in the crystalline organic semiconductor blends using ultraviolet photoemission spectroscopy, optical characterization and theoretical simulations. In particular, the tuning of energy structure in the blend crystalline films was attributed to the direct effect of tight-binding Hamiltonian arising from molecules with different electronic properties. We believe our results would stimulate further work on the growth and epitaxy of molecules with different chemical structures and tuning of their band structure properties for optoelectronic device applications.

**Experimental Section**

**Sample preparation.** Silicon with native oxide and glass wafers were first cleaned with acetone, isopropanol under ultrasonication followed oxygen plasma cleaning before loading to a Lesker vacuum deposition system (base pressure < $10^{-7}$ mbar). For blend film seed layer crystallization, molecules were co-evaporated with individual quartz crystal monitor to adjust the blend ratio. The film thickness was kept at 40 nm. For the samples prepared by epitaxial growth, orthorhombic RUB crystals was first prepared by annealing a 5 nm Di-[4-(N,N-ditolyl-amino)-phenyl]cyclohexane (TAPC)/40nm RUB bilayer at 160 °C for 4 minutes as reported previously[22] followed by the deposition of the blend film and heterostructures with a deposition rate of 0.6 nm/s. The charge transport devices were prepared in the Lesker system by deposition through shadow masks in sequence forming cross-bar devices with different active area sizes 0.2, 0.5, 1 and 2 mm square.

**Measurements.** Polarized optical images were taken with a Nikon Eclipse LC100 PL/DS polarization microscope. Atomic force microscope (AFM) images were taken with an AIST-NT Combiscope1000. The optical absorption measurements were performed with a Shimadzu UV-3100 UV-VIS-NIR spectrophotometer. The emission spectra were measured with a fluorescence spectrometer (Edinburgh Instruments). The GIWAXS investigation was performed at the ID10 beamline of the ESRF synchrotron in Grenoble, France. An area detector (Pilatus 300k), which was placed approximately 16 cm behind the sample, was used to record the images. The exposure time was between 0.3 s and 1 s. The beam size was 40 μm horizontally and 40 μm vertically and the beam energy was 10 keV. The incidence angle was 0.1432°. The measurements were calibrated using a $LaB_6$ scattering standard and were analyzed with the WxDiff software (by S.C.B.M.). The samples were transferred to an ultrahigh vacuum chamber (ESCALAB 250Xi by Thermo Scientific, base pressure: $2 \times 10^{-10}$ mbar) for the UPS measurements. The measurements were then carried out using a He discharge lamp (hv = 21.2 eV) and a pass energy of 2 eV.

**Acknowledgements**

S.-J.W acknowledge funding from DFG Project, WA 4719/2-1. We acknowledge support from the DFG Koselleck project No. 456344071. We acknowledge the European Synchrotron Radiation Facility (ESRF) for provision of synchrotron radiation facilities and we would like to thank Maciej Jankowski and Oleg Konovalov for assistance and support in using beamline ID10. F. T. and S. M. acknowledge financial support from the German Research Foundation (DFG, MA 3342/6-1) and the Graduate Academy of TU Dresden. H.K. acknowledges financial support from the German Research Foundation (KL 2961/8-1). This project has received funding from the European Research Council (ERC) under the European Union's Horizon 2020 research and innovation programme (ERC Grant Agreement n° 714067, ENERGYMAPS). We thank Andreas Wendel and Tobias Günther in IAPP for their help to prepare the samples.

**Figures**

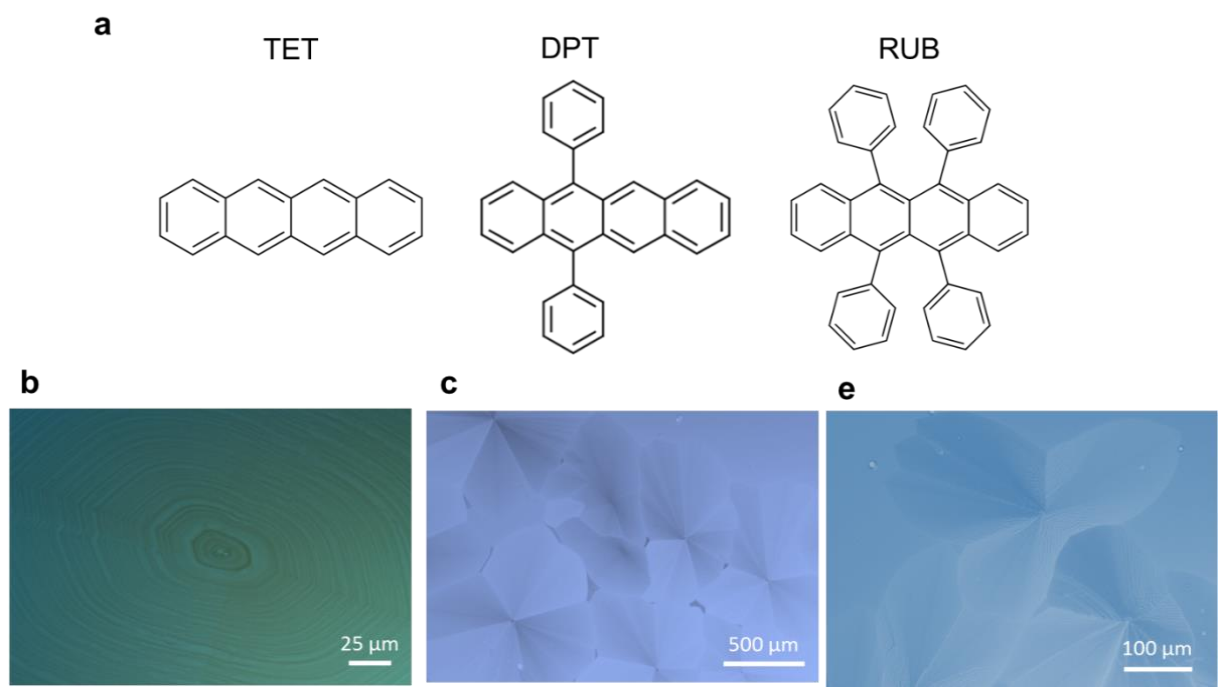

**Fig. 1. Molecular structures and polarized microscope images for the materials investigated.** (**a**) Chemical structures of tetracene (TET), 5,12-Diphenyltetracene (DPT) and rubrene (RUB). Polarized optical image of DPT:RUB blend crystals with ratio 1:10 (**b**), 1:3 (**c**) and 1:2 (**d**). The thickness of the blend films was 40 nm.

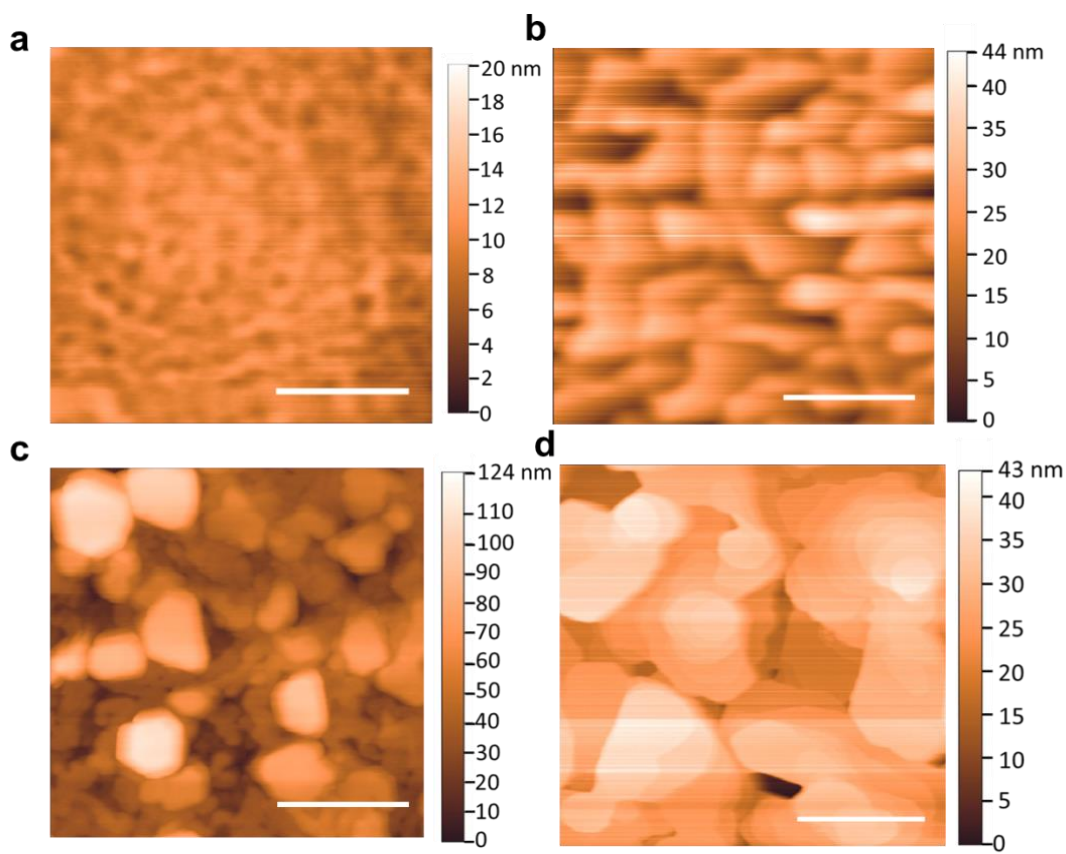

**Fig. 2. Surface topography of the blends and heterostructures.** Atomic force microscope images of DPT:RUB blend with 1:1 ratio (**a**), DPT film (**b**), TET:RUB blend with 1:1 ratio (**c**) and TET film (**d**). All layers are deposited on top of a 40 nm thick orthorhombic rubrene thin film crystals. The scale bars denote 1 micron.

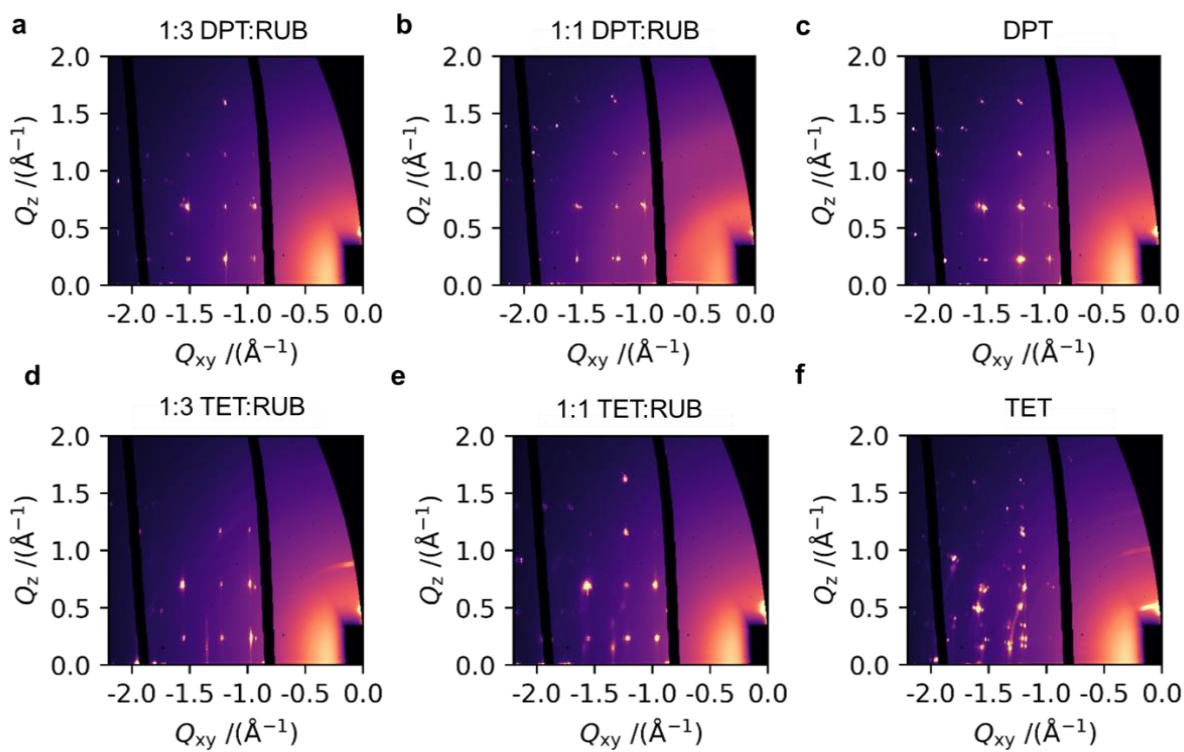

**Fig. 3. X-ray diffraction of the blends and heterostructures.** GIWAXS diffraction patterns of DPT:RUB blend with 1:3 ratio (**a**), DPT:RUB blend with 1:1 ratio (**b**), DPT film (**c**), TET:RUB blend with 1:3 ratio (**d**), TET:RUB blend with 1:1 ratio (**e**) and TET film (**f**). All layers are deposited on top of a 40 nm thick orthorhombic rubrene thin film crystals.
15

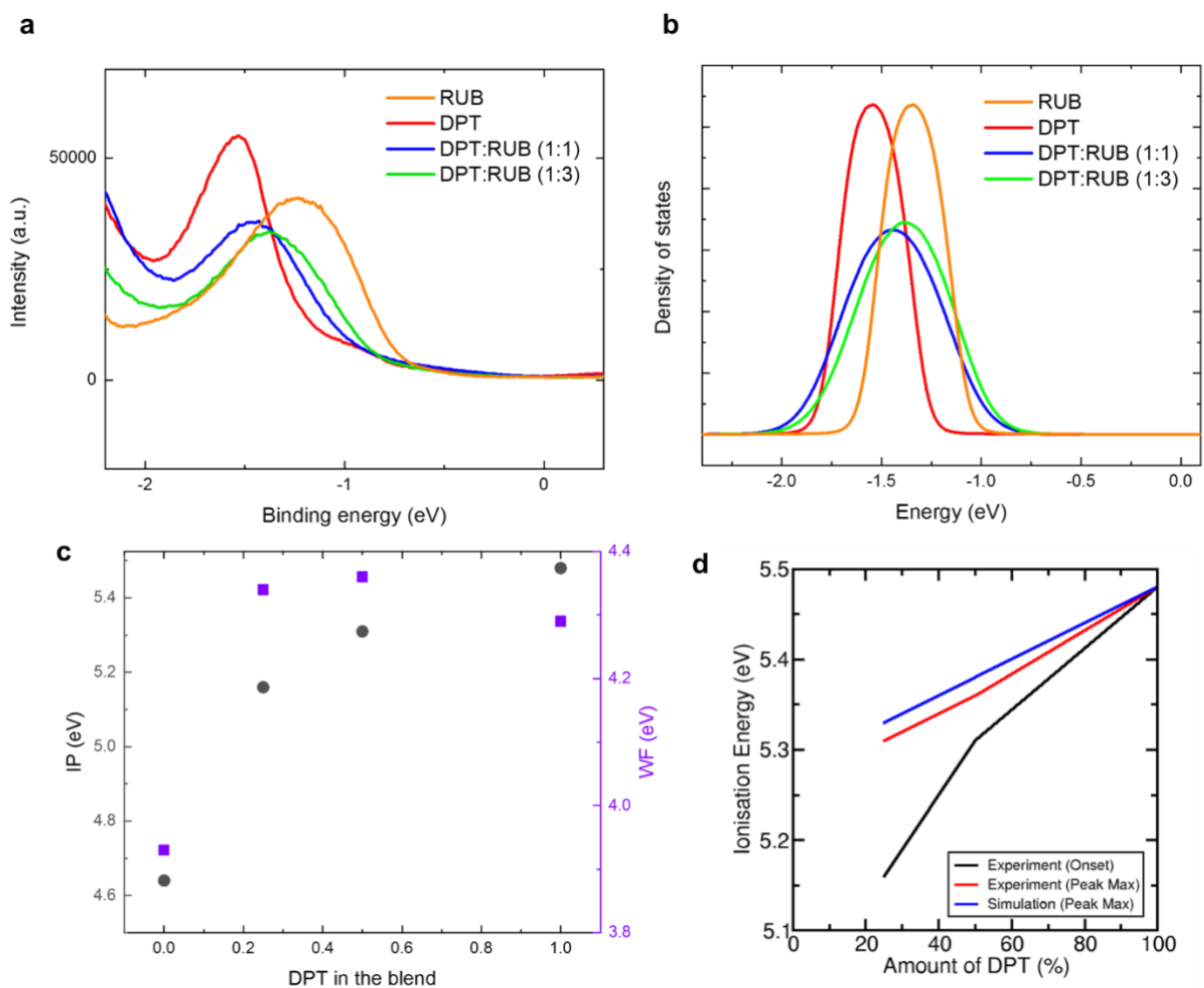

**Fig. 4. Ultraviolet photoelectron spectroscopy and theoretical simulation results.** (**a**) UPS spectra of RUB:DPT blend films. (**b**) Electronic density of states in dependence of the blending ratio for effective disordered system. There is a shift of the peak maximum in due to the varying composition of DPT and RUB. The shift of the maximum peak position is compared to experimental values in Fig. 4d. (**c**) Ionization potential and work function of DPT:RUB blend with various mixing ratio. The ionization potential is defined by the edge of the HOMO energy level (Fig. S6a). (**d**) Comparison between experimental ionization potential estimated from the peak maxima and onset and simulation results based on the peak maxima.

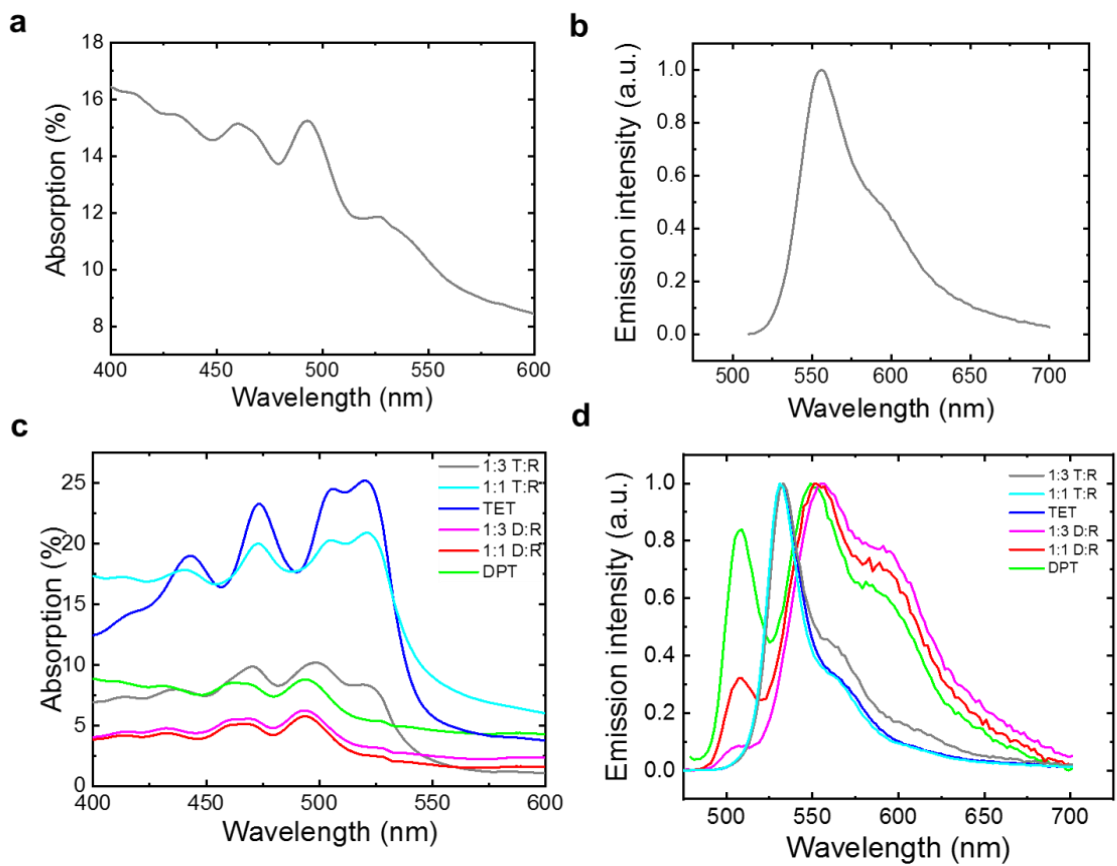

**Fig. 5. Absorption and emission of the blend films.** Absorption (**a**) and emission (**b**) spectrum of a 40 nm thick orthorhombic RUB film. Absorption (**c**) and emission (**d**) spectra of blend films with a 40 nm thick orthorhombic RUB seed layer. The excitation wavelength for the emission measurement was kept around 450 nm.

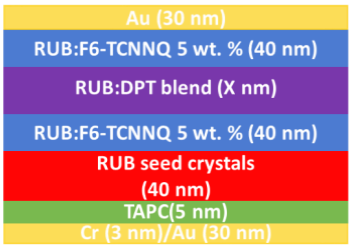 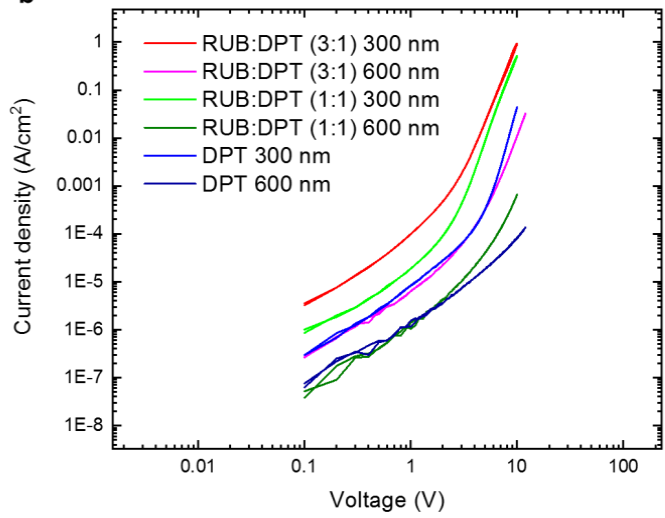

**Fig. 6. Charge transport of the blend films.** (**a**) Layer stack of the hole-only diode used to investigate the charge transport of the blend films. (**b**) Current density-Voltage (*J-V*) characteristics of the blend films with different ratios and thicknesses.